# Multivariate Ensemble Forecast Framework for Demand Prediction of Anomalous Days

Muhammad Qamar Raza, *Student Member, IEEE,* N. Mithulananthan, *Senior Member, IEEE,* Jiaming Li, Kwang Y. Lee, *Fellow, IEEE*

***Abstract*--An accurate load forecast is always important for the power industry and energy players as it enables stakeholders to make critical decisions. In addition, its importance is further increased with growing uncertainties in the generation sector due to the high penetration of renewable energy and the introduction of demand side management strategies. An incremental improvement in grid-level demand forecast of anomalous days can potentially save millions of dollars. However, due to an increasing penetration of renewable energy resources and their dependency on several meteorological and exogenous variables, accurate load forecasting of anomalous days has now become very challenging. To improve the prediction accuracy of the load forecasting, an ensemble forecast framework (ENFF) is proposed with a systematic combination of three multiple predictors, namely Elman neural network (ELM), feedforward neural network (FNN) and radial basis function (RBF) neural network. These predictors are trained using global particle swarm optimization (GPSO) to improve their prediction capability in the ENFF. The outputs of individual predictors are combined using a trim aggregation technique by removing forecasting anomalies. Real recorded data of New England ISO grid is used for training and testing of the ENFF for anomalous days. The forecast results of the proposed ENFF indicate a significant improvement in prediction accuracy in comparison to autoregressive integrated moving average (ARIMA) and back-propagation neural networks (BPNN) based benchmark models.**

***Index Terms*--Short term load forecasting (STLF), ensemble forecast framework (ENFF), Elman neural network (ELM), feedforward neural network (FNN), radial basis function (RBF), global best particle swarm optimization (GPSO), trim aggregation.**

## I. Introduction

Load forecasting is extremely important for energy planning and security to provide uninterruptible power supply to the consumers in the cheapest possible way [1]. A number of power system operations including contract evaluation, scheduling, planning, adjustment of tariff rates and maintenance can be carried out based on load forecasting [2]. Load demand forecasting is always important in power industry planners and policy makers as it enables stakeholders to make critical decisions. An accurate load forecasting can also be helpful for efficient energy management. In the last decade, extensive research has been published on load forecasting due to its potential for demand side management and smart power grids [3].

Muhammad Qamar Raza and N. Mithulananthan are with the Power and Energy System group, at School of Information Technology and Electrical Engineering, University of Queensland, Brisbane, QLD 4072, Australia (e-mail: qamar.raza@uq.edu.au; mithulan@itee.uq.edu.au).
K. Y. Lee is with Department of Electrical & Computer Engineering, Baylor University, Waco, TX 76706 USA (e-mail: Kwang_Y_Lee@baylor.edu).
Jiaming Li is associated with Data61, Commonwealth Scientific and Industrial Research Organization (CSIRO), Sydney, Australia (jiaming.li@csiro.au).

Accurate load forecast of especially anomalous days becomes more crucial as overall power system efficiency can be affected due to under- and/or over-estimation of the load. Power supply and demand response can be severely affected by under-estimation of demand. In addition, due to the unavailability of large backup supply, overload conditions can create power system stability issues. On the other side, over-estimation of load demand can cause a rise in the production cost [4]. Power system operation, planning and maintenance along with other critical decision making for utility and prospective regulator can be carried out based on accurate load forecasting [5].

An accurate load forecasting of anomalous days is a challenging task as multiple meteorological and exogenous variables are affecting the load profile. There is a strong impact of weather variables on load demand, (such as temperature, relative humidity, dew point, dry bulb, wind speed, cloud cover, and the human body index). The multiple loads and consumer preferences also create an enormous impact on load demand. To accurately forecast demand especially anomalous load demand, correlated exogenous and meteorological variables need to be considered as the input to the forecasting framework. In addition, it is also very crucial to select better performing models and their proper training with highly correlated input for high prediction accuracy. It is also necessary to have enough correlated training samples for accurate load forecasting. For load forecasting of a normal day (a day with a predictable load profile), the training data have enough correlated training samples to train the network. However, an anomalous day (irregular or out of routine days) load forecasting has a much smaller number of patterns for effective training of the network. Therefore, the anomalous day forecasting model is more complex and difficult to design for higher forecast accuracy [6]. Generally, the prediction accuracy of models for anomalous days is lower due to multiple factors such as uncertainty in demand, meteorological variables, unpredictable sociological events and intermittency of renewable energy resources, etc.

Anomalous days include all irregular days such as public holidays, country social or cultural events, preceding or following days of holidays and special events [7, 8]. The frequency of anomalous days is less than that of normal routine days, but it negatively affects overall power system reliability and energy security. There are much fewer data patterns for network training to predict anomalous days' load demand. Therefore, to design an accurate anomalous day forecasting model, data length has to be longer than the usual for proper training of the network (e.g., a five-year historical load data having similar characteristics of a load profile).

A large number of studies have been performed on short term load forecasting (STLF) in the past due to its huge impact on the economy. The hybrid Kalman filters [9], autoregressive models [11], ARMAX model [10, 11], statistical techniques [12] and composite modeling [13] are also applied. Some of the techniques are not robust enough to deal with uncertain meteorological and sociological events. Consequently, these parametric techniques produce high forecast errors under uncertainty conditions, which is a major drawback. Artificial neural network (ANN) techniques also gained attention from researchers due to its capability for a range of forecast applications [14]. Several techniques have been applied to STLF, such as support vector regression techniques [15], artificial immune system [16], radial basis function [17], Hybrid Monte Carlo algorithm based ANN forecast model [18] and feedforward neural networks [19-21]. Some of the forecasting models achieved the good forecasting accuracy but to a certain level. In [22], a multi-objective genetic algorithm based fuzzy classifier was designed to classify the anomalous days with the overall aim to improve the prediction accuracy. Proposed method improved prediction accuracy in some of the selected days but unable to provide stable and improved forecast results. It is worthwhile to mention that, very few researches have been published on anomalous or special days forecast. Most of the existing models are unable to deal with the uncertainty due to meteorological conditions and other variables affecting load demand.

A single model can forecast load demand up to a certain degree of accuracy due to the limitations of a particular model. The forecast accuracy can be improved by overcoming the inadequacy of a single and hybrid model by designing a framework which does not affect the prediction performance of individual predictors. Overall prediction accuracy of ensemble framework even with low performing predictors can be increased in forecast applications [23, 24]. In our previous work an ensemble framework was proposed to predict the building level load demand [24]. The proposed framework shows a significant improvement in comparison with benchmark models by undertaking the incapability of single predictor or models. The diverse output of each independent predictor provides an opportunity to enhance the overall forecast by exploring all viable solutions. Multiple predictors organized in a systematic way called ensemble network can provide better prediction results as compared to a single predictor. Therefore, an ensemble forecast framework (ENFF) is proposed with a systematic combination of three different predictors. In this research Elman neural network (ELM), feedforward neural network (FNN) and radial basis function (RBF) neural network were selected due to their individual predictor performance for forecasting application with highly fluctuating training data and predict PV output power and other applications [23]. In addition, a number of test case studies was designed to assess to the performance of ELM, FNN and RBF for anomalous days forecast. The major contributions of the proposed novel ensemble approach are highlighted as follows:

1. Identification of strongly correlated meteorological and exogenous variables for the training of neural predictors in ENFF using autocorrelation function (ACF).
2. Development and integration of the three different predictors, i.e., Elman neural network (ELM), feedforward neural network (FNN) and radial basis function (RBF) neural network in ENFF.
3. Training of neural predictors using global particle swarm optimization (GPSO) for higher forecast accuracy.
4. Aggregation of neural predictors output using trim aggregation technique by removing forecast error extremes.
5. Significant improvement in prediction accuracy of the proposed framework for anomalous days in comparison with benchmark forecasting models.

Remaining paper is organized as follows: Section II describes the forecast inputs of predictors, topology and data observation. Section III describes the working of ensemble forecast framework, PSO algorithm, training of neural network with the GPSO algorithm, and the trim aggregation technique. Section IV provides the results of anomalous days forecast using the proposed model, and the comparison with benchmark forecasting models for validation.

## II. Ensemble Framework Topology and Inputs

### *A. Data Preparation*

Six years of data, from 2004 to 2009, is collected from New England ISO for training and validation of the framework. The collected data contains load demand along with meteorological variables (dry bulb and dew point), which is used to predict the future load demand [25]. Six-year hourly load form January 1, 2004, to December 31, 2009, is shown in Fig. 1. From the load profile analysis, a seasonal trend can be observed in the load data series. A repeated cycle of load demand is recorded in different seasons of the year due to change in meteorological conditions. A seasonality trend can be observed with peak and low load demand with seasonal temperature variations.

### *B. Correlated Forecast Framework Inputs and Training Data*

Input data for ENFF is divided into different data sets for training and testing. Network training of neural predictors is performed using the training data set. To measure the performance of the forecasting framework, the testing data set is utilized. In literature, the year 2008 hourly load and weather data of the New England ISO grid is used to train the neural network [24]. However, a larger data set is used to train the network due to the lack of correlated data samples in the one-year data set. Anomalous-day load data for the year 2009 is used for testing and measuring the accuracy of the forecasting framework. Selection of correlated input variables is very crucial for better training and accuracy of the prediction models. There are no specific rules or criteria for input selection of the forecasting framework, but a suitable selection can be carried out based on the expertise and field experience [26].

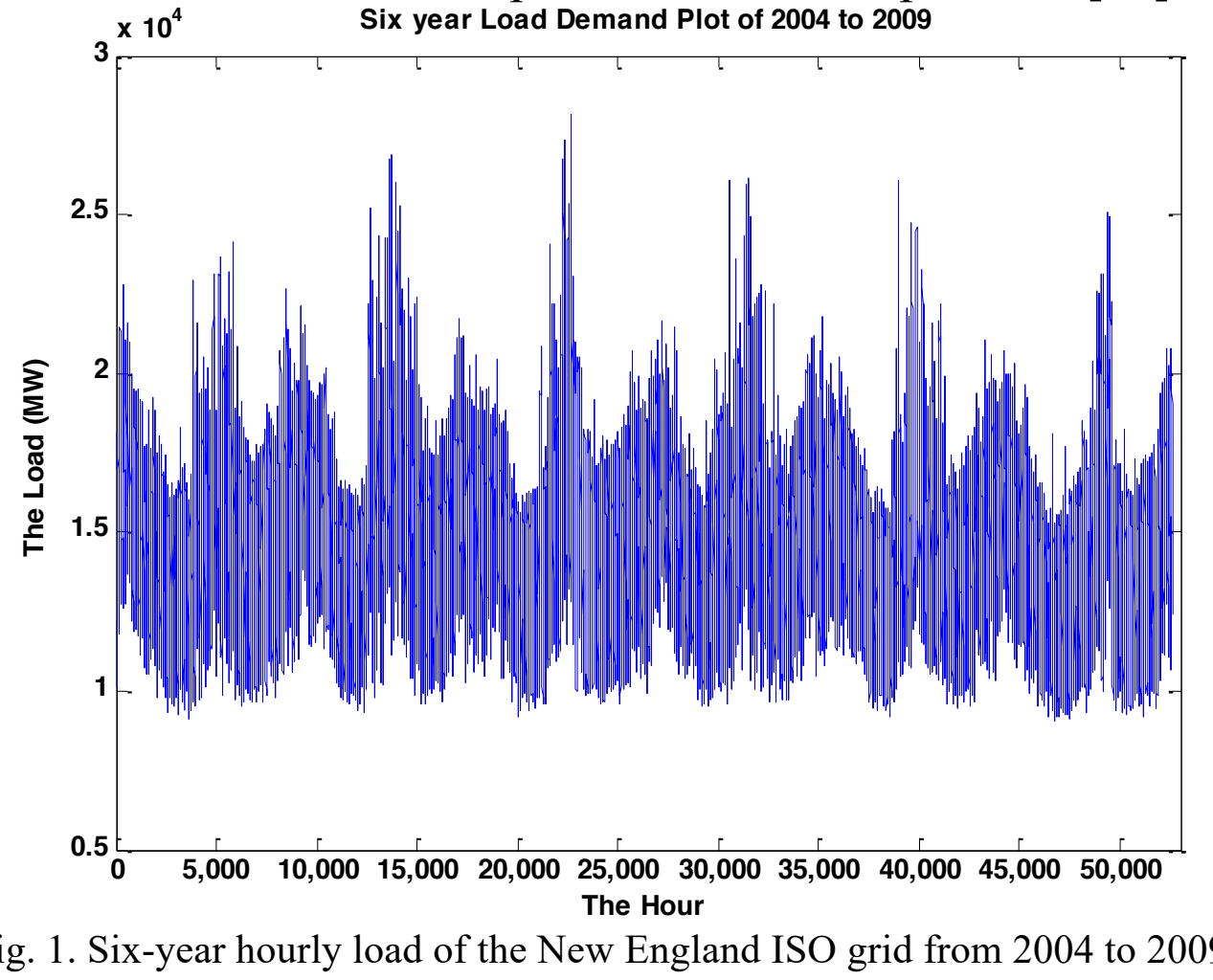

Fig. 1. Six-year hourly load of the New England ISO grid from 2004 to 2009.

In this paper, highly correlated variables (which has an impact on demand) are selected for the forecasting framework based on correlation analysis. The selected correlated inputs are shown in Fig. 2, where

1. $L_d(w, d, h)$ denotes the load demand of a particular hour of the day of the week;
2. $L_d(w, d, h\text{-}1)$ represents the load demand of the previous hour of the same day of the same week;
3. $L_d(w, d\text{-}1, h)$ exemplifies the load demand of the same hour of the previous day of the same week;
4. $L_d(w\text{-}1, d, h)$ is the load demand of the same hour of the same day of the previous week;
5. Day of the week (signifies the day of the week such as a first or fifth day);
6. Day type (Working day or off day);
7. Hour of the day;
8. Dew point temperature;
9. Dry bulb temperature.

It is observed that the load demand during work and off days are different due to variation in human activities. Therefore, the type of the day is considered as an input of the forecast framework to reflect the human activities on load demand. The demand also varies during different times of the day and days of the week. Load profile analysis depicts that demand varies throughout the day and days of the week. The change in load demand varies according to an hour of the day and day of the week. A similar and recurring cycle can be observed in a one-month load profile of demand with respect to hours of the day and days of the month. In addition, an analysis of load profile also indicates that the load demand in weekends is quite different and more uncertain in comparison with weekdays. To design and develop an accurate forecast framework, the type and hour of the day are also included as inputs of the forecast framework.

### C. *Weather Input Variables*

Several studies have been conducted in the past to investigate the impact of meteorological conditions on load demand [26, 27]. Fig. 3 presents the relationship between the dry bulb temperature and load demand. The graph shows a nonlinear relationship between the dry bulb temperature and load demand. Human perception studies show that the dry bulb temperature in the range of 45°F to 65°F is suitable for human living. The load demand is low within this range of dry bulb temperature.

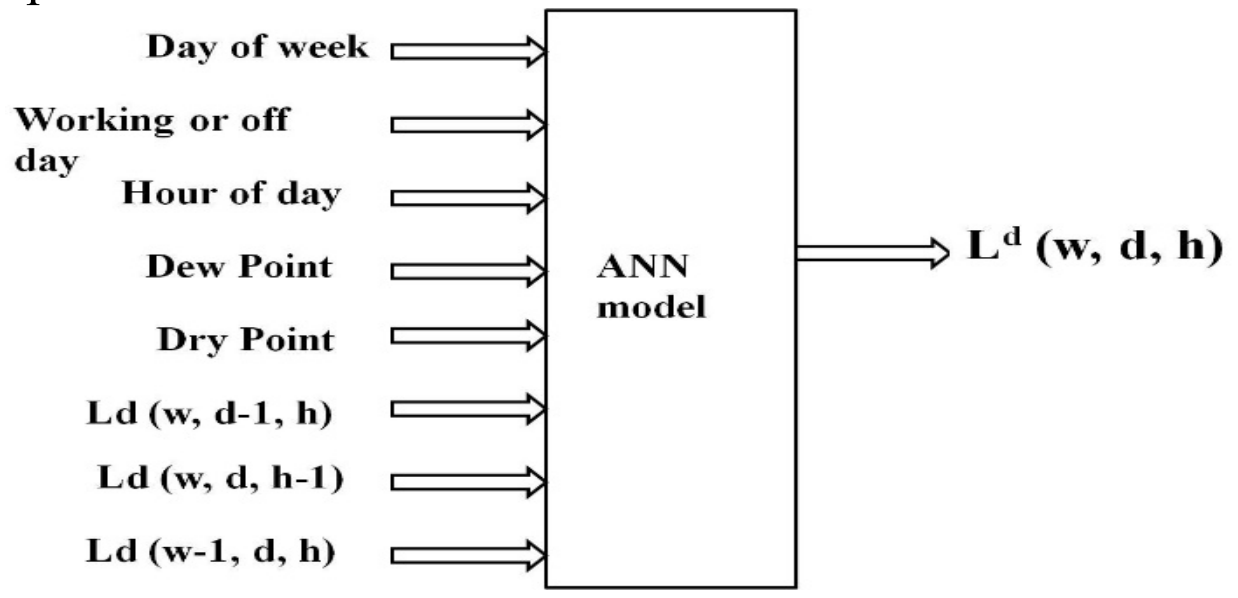

Fig. 2. Selected inputs of Multivariate ENFF's predictor [28].

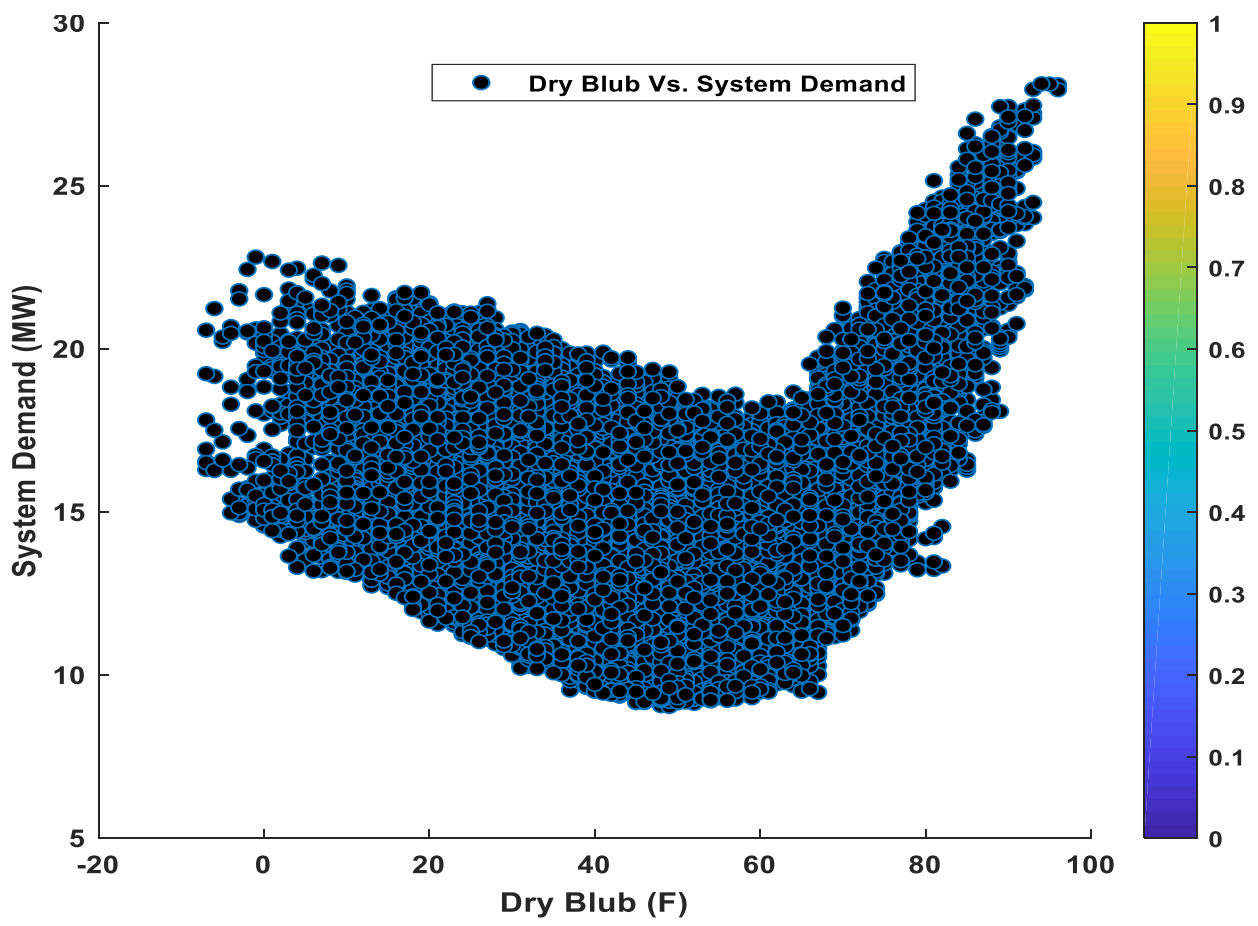

Fig. 3: Relationship dry bulb vs. power load demand.

## III. Proposed Forecast Framework

An ensemble forecast framework (ENFF) is proposed, which consists of three different predictors, namely Elman neural network (ELM), feedforward neural network (FNN) and radial basis function (RBF) neural network. These neural predictors are selected based on their individual performance for different forecast applications [23]. The proposed neural network ensemble (NNE) framework is shown in Fig. 4. The working of neural predictors in multivariate NNE can be described in five different steps as given below.

*Step 1 (Data Preprocessing):* The real time recorded data of New England ISO consist of different meteorological variables along with load demand. By using correlation analysis, correlated inputs are selected to train the ENFF. Therefore, historical correlated load demand and meteorological variables are applied for training.

*Step 2 (Wavelet Transform):* To train the predictors in the ensemble framework effectively, there is a need to remove the fluctuations and spikes in historical load demand. Therefore, wavelet transformation (WT) is used, which decomposes the time domain data and approximates the components. These decomposed signal shows relatively much more stable behavior and improves the training of predictors. Therefore, the historical data set is processed with WT. The WT can be classified into two groups such as continuous WT (CWT) and discrete WT (DWT). The CWT component of WT can be defined as [23]

$$CWT_x(a,b) = \frac{1}{\sqrt{|a|}} \int_{-\infty}^{+\infty} \Psi^*(t) x(t)\,dt, \quad a > 0 \tag{1}$$

$$\Psi_{a,b}(t) = \frac{1}{\sqrt{|a|}} \Psi^*\left(t - b/a\right), \quad a > 0, -\infty < b < +\infty \tag{2}$$

where $x(t)$, $\Psi_{a,b}(t)$, $a$, $b$ and $*$ are original input, mother wavelet signal, scaling factor, a parameter for translating and conjugate complex parameter, respectively. The DWT can be completed by discretized translating and scaling of the mother signal as given in [29]:

$$DWT_x(m,n) = 2^{-(m/2)} \sum_{t=0}^{T-1} x(t) \Psi\left(t - n2^m / 2^m\right) \tag{3}$$

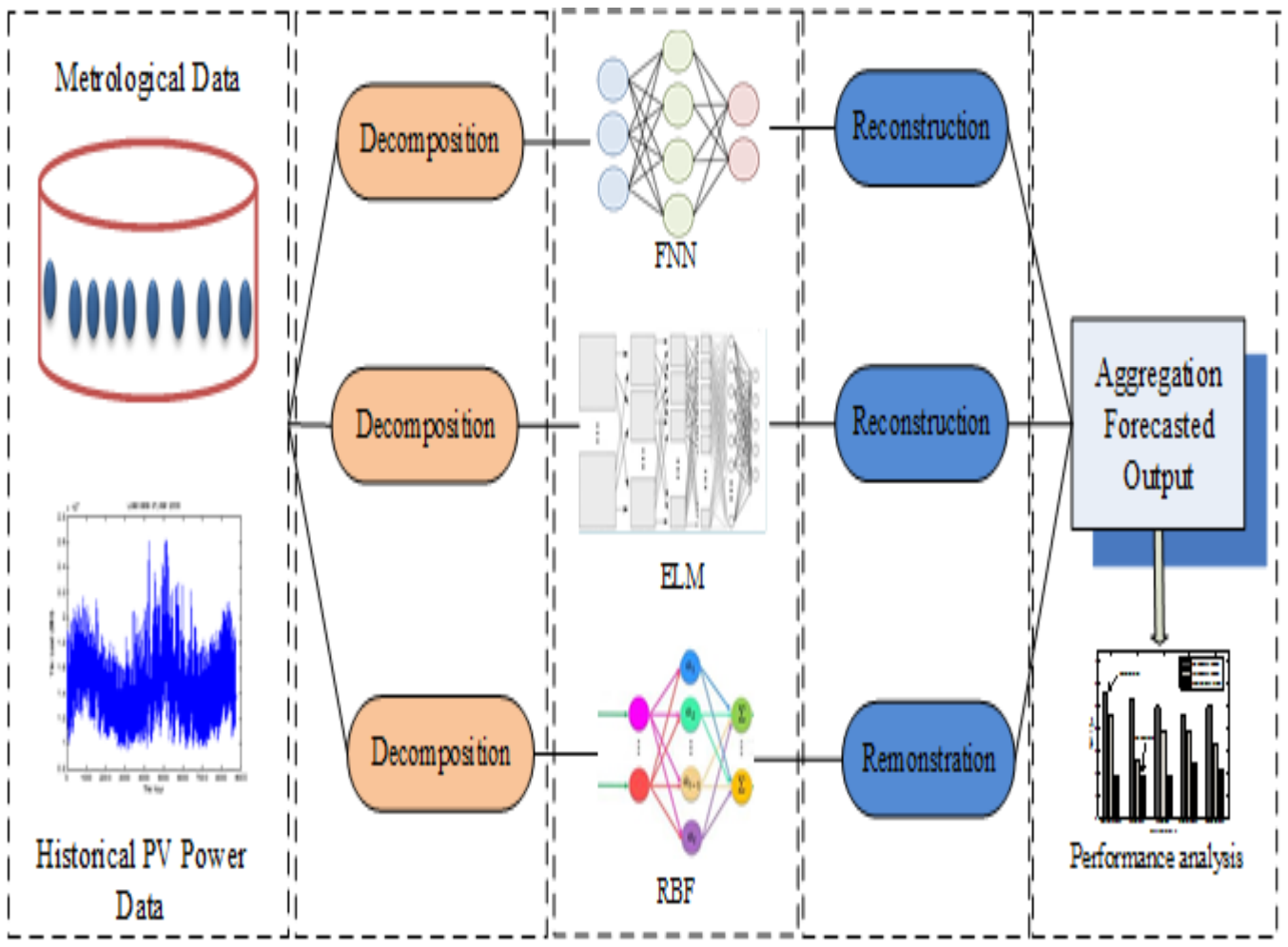

Fig. 4: Proposed Neural based Ensemble forecast framework.

which implies in (2) the signal $x(t)$, $t$ is time index, $a = 2^m$ and $b = n2^m$. Mallat's multiresolution technique is used for pre-processing of input [30]. Using high and low pass filters, the original load demand signal is transformed into detailed and approximation components. Then a three-level WT decomposition was applied. During the decomposition process of WT, high pass (H.P.) filter generates the high frequency or detailed components of the signal. The output of predictor is reconstructed using reconstruction process.

Step 3 (Construction of Predictors): The ENFF is designed by using ELM, FNN and RBF. In addition, ELM, RBF and FNN are trained using global particle swarm optimization (GPSO) to enhance the individual predictor's performance. As a result, it will lead to higher forecasting performance. The details of NN training with GPSO can be found in [31]. GPSO algorithm is described in Section III-C. It is worth mentioning that different predictors were used to have forecasting diversity. The diverse forecasting results will lead to enhanced overall forecasting accuracy by exploring multiple solutions.

Step 4 (WT reconstruction): Forecasted output of the neural predictor is reconstructed using the WT reconstruction process. The components are combined to generate the output.

Step 5 (Output Combination): The output of each predictor in the ensemble network is combined using a combination or aggregation method. In this research, a trim aggregation technique is used to combine the output of each predictor by removing forecast error extremes.

### A. Trim Aggregation to Combine the Output

Each predictor in the proposed ENFF will produce a forecasted output according to its individual performance and performance affecting parameters. The forecast output of NN predictors may differ from each other due to differences in NN architecture and network initialization parameters. Some of predictor's generate forecast outliers, which is certainly away from test data set. In order to get the accurate ensemble network forecast output, the output of each predictor is combined using an aggregation technique after removing the outliers. Discarding the worst performing model output using a trimming parameter $\alpha$ is an effective way to remove the lower and upper extreme forecast. Although there is no exact method available to determine the value of the trimming parameter $\alpha$, a methodology to determine the optimal value of the trimming parameter $\alpha$ is proposed [32]. It is assumed that $\alpha \in [0,100]$. The trimming amount can be determined using the network validation data by sorting them in ascending order. Then the trimming number can be calculated using the trimming amount of NN ensemble output. The output trimming can be defined by [32]:

$$NN_{trim} = \frac{\alpha}{100} NN_{total} X_{\alpha} \quad (4)$$

where, $X_{\alpha}$ is defined as in Eq. 5.

$$X_{\alpha} = \left[ \frac{NN_{trim}}{2} + 1, \frac{NN_{trim}}{2} + 2, ..., N_{total} - \frac{NN_{trim}}{2} \right] \quad (5)$$

The mean of $\alpha$ trimmed forecast output for all forecasts can be calculated as given in Eq. 6.

$$\hat{Z}_i^{\alpha} = \frac{1}{NN_{total} - NN_{trim}} \sum_{j \in I_{\alpha}} \hat{Z}_{i,j} \quad (6)$$

Where $\hat{Z}_i^{\alpha}$ is trimmed output with $\alpha$ trimming parameter, $D_{vald}$ validation data and $i = 1,2,...,D_{vald}$. The forecast error of $\alpha$ trimmed can be calculated as $MAPE^{\alpha}$ given in Eq. 7.

$$MAPE^{\alpha} = \frac{1}{N_h} \sum_{i=1}^{N_h} \left| \frac{\hat{Z}_i^{\alpha} - Z_i}{Z_i} \right| \quad (7)$$

where $N_h$ is forecast horizon, $Z_i$ is desired output at the time $i$ and $\hat{Z}_i^{\alpha}$ is the trimmed output. In order to get the accurate forecast output, the output of each predictor is combined using an aggregation technique after removing the outliers.

### B. Global Best Particle Swarm Optimization (GPSO)

Particle swarm optimization (PSO) is a heuristic search approach based on the population of particles in a swarm. The PSO shows a good capability to find an optimal solution to a wide range of real world problems [33]. Further research leads to the Global best or Gbest technique, which is a modified version of the PSO technique. It tries to achieve the global optimum solution by updating particle positions and velocity in a swam [34]. In GPSO, the new particle position is influenced not only by its own previously visited best positions but also by the best of the position of its neighborhood particles. The GPSO finds the best solution for all particles in the swarm and has a higher probability to achieve the global optimum solution. Therefore, in the position update process it reflects the social influence of all particles as given below in Eqs. 8 and 9 [35]:

$$v_i^{(k+1)} = wv_i^k(t) + c_1 r_1 (y_i^k(t) - x_i^k(t)) + c_2 r_2(t)(y_i^k(t) - x_i^k(t) \quad (8)$$

$$x_i^{(k+1)} = x_i^k(t) + v_i^{(k+1)}(t) \quad (9)$$

where:

$c_1$ and $c_2$: constants used to define the contribution of the social and cognitive component.

$r_1$ and $r_2$: the randomly generated numbers in the range [0, 1];

$v_i^k$ : velocity of the $i$-th particle in dimension $k$ direction at time $t$.

$x_i^k$ : position of the $i$-th particle in dimension $k$ direction at time $t$.

### C. Neural Network Training using GPSO

The training process of neural predictors with GPSO is based on the position of the current particle in a swarm. Each particle in a swam characterizes the potential solution to reduce the training of neural predictor to a threshold level. Neural network weights are represented by particle positions. Dimensionality of a particle in the search space represents the associated number of NN weights. In each training epoch, particles try to update their velocity and positions based on NN training error as the objective function. Particle tries to achieve a global optimum solution by updating particles position and velocity according to (8) and (9). This iterative learning process is continued until the threshold of learning error is achieved. In this way minimum network training is achieved as the swam is at the global optimum point. Fig. 5 represents the learning steps of a neural network with PSO [28]:

1. Swam population is initialized to start the learning process of the network with constrained random position and velocity. In addition, these initial particle position is allocated as network training parameters (weight, bias).
2. At each training iteration, the network is trained to forecast the load demand.
3. Learning error (the difference between target and current output) is generated by the Network in the epoch of training process.
4. To minimize the learning of the network, particle positions (weight and bias) are updated. New particle velocity is calculated based on "Pbest" of each particle in a swarm and "Gbest" of the swarm to reduce the training error according to (8).
5. New particle positions are calculated based on new velocity and previous position according to (9). These new positions are assigned to network training parameters to new learning error in next iteration.

This training process of the network is continued until the termination criteria of network learning error or number of iterations are met.

## IV. Results and Discussion

Mean absolute percentage error (MAPE) is calculated for each anomalous day case study to assess the performance of the proposed framework as given below:

$$\text{MAPE} = \frac{1}{M}\sum_{i=1}^{M}\frac{\left|L_{actual} - L_{predicted}\right|}{L_{actual}} \tag{7}$$

where $L_{actual}$ and $L_{predicted}$ are the actual and the forecasted load, respectively, and $M$ is the number of data points. To analyze the performance of the proposed forecast framework, case studies for anomalous days are designed. The selected anomalous days are Christmas Day, (Friday, December 25, 2009), Memorial Day (Monday, May 25, 2009), Easter Day (Sunday, April 12, 2009), Labor Day (Monday, September 7, 2009) and New Year's Day (Thursday, January 1, 2009). The BPNN and ARIMA forecasts are also applied to predict the anomalous day's load demand and compared with the proposed ENFF technique. Figs. 6-10 show the 24-hour ahead load forecast of Christmas Day, Memorial Day, Easter Day, Labor Day and New Year's Day, respectively, using all aforementioned techniques. To forecast the load demand of anomalous days, the neural network architectures are designed varying network architecture to achieve the diversity and explore forecast solutions. With Core I7 processing machine, 8GB RAM and DDR storage, each training iteration is completed in 127 seconds or less time. In addition Benchmark models are provided below.

### A. Backpropagation Neural Network (BPNN)

In the last decade, BPNN is applied to a vast range of forecast applications such as electrical load, price and wind forecast with a good level of success. In BPNN, backpropagation process tries to determine the nodes connections weights values. The weights values of the network are updated using error function. The network learning error is calculated with a difference of network predicted and actual. The error between the predicted and actual output values is back-propagated and weight values are updated based on network learning error. Therefore, the network learning error is minimized by back propagating the error and updating the connections weight values.

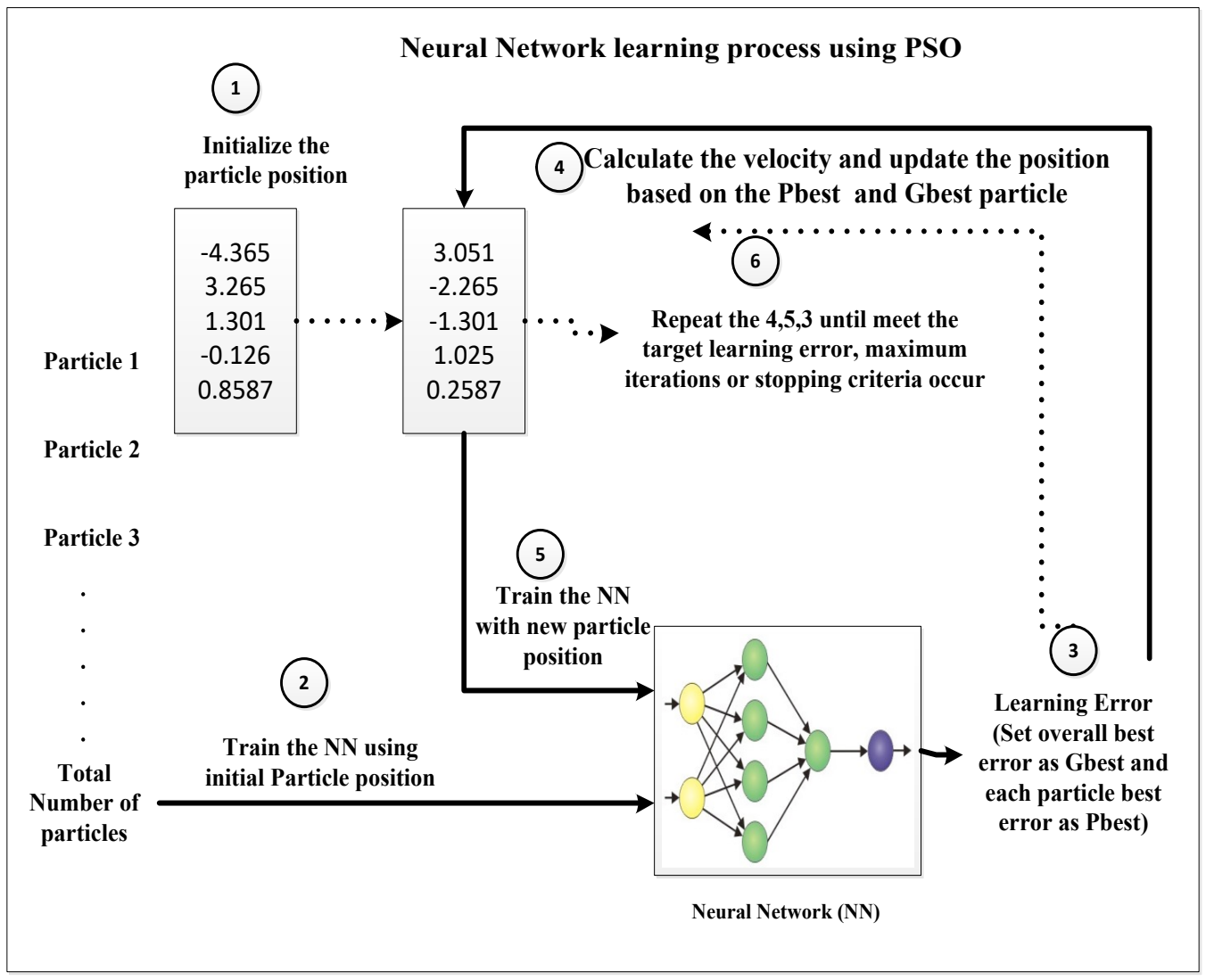

Fig. 5. NN learning process using PSO algorithm [28].

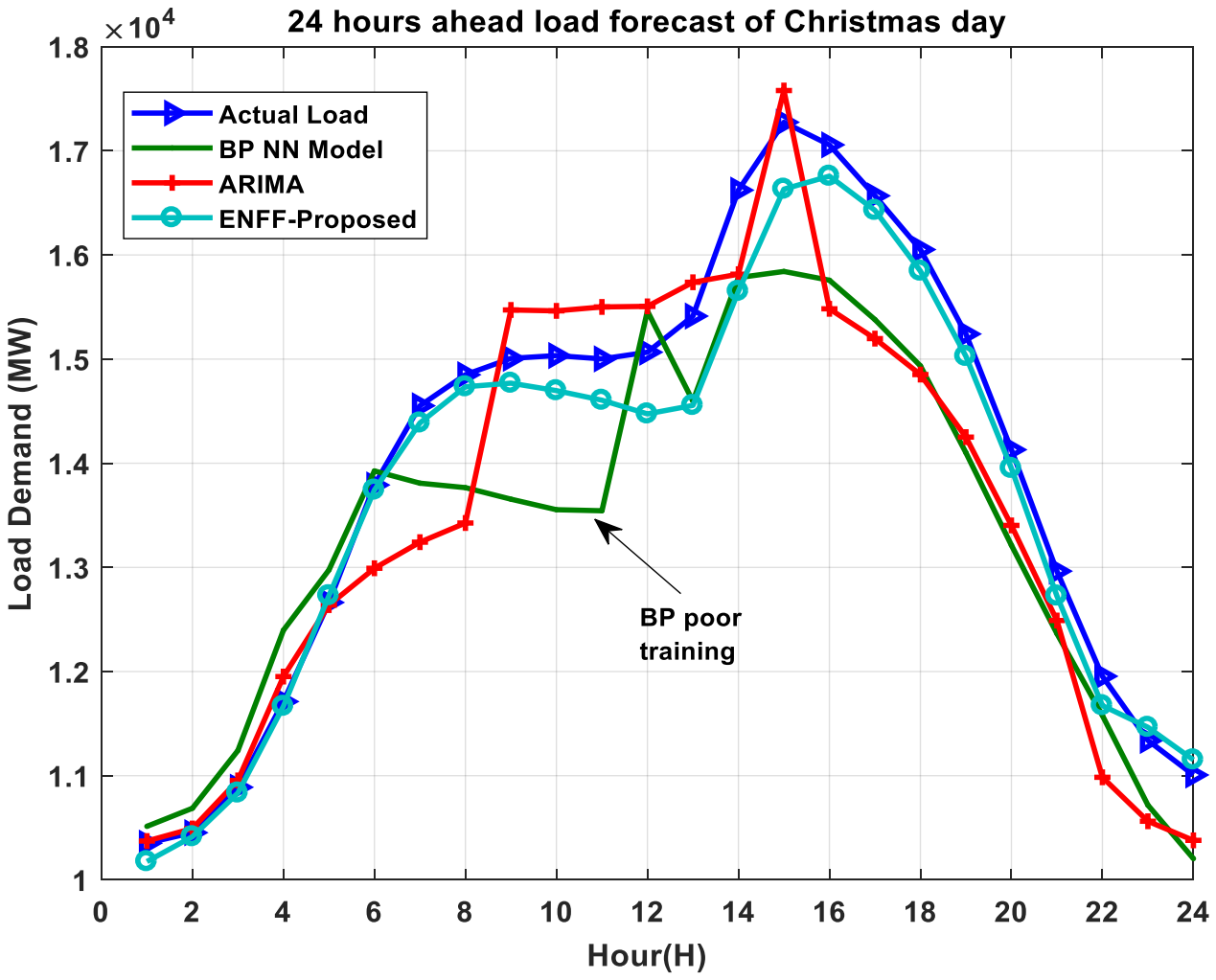

Fig. 6. The 24-hours ahead load forecast of Christmas Day.

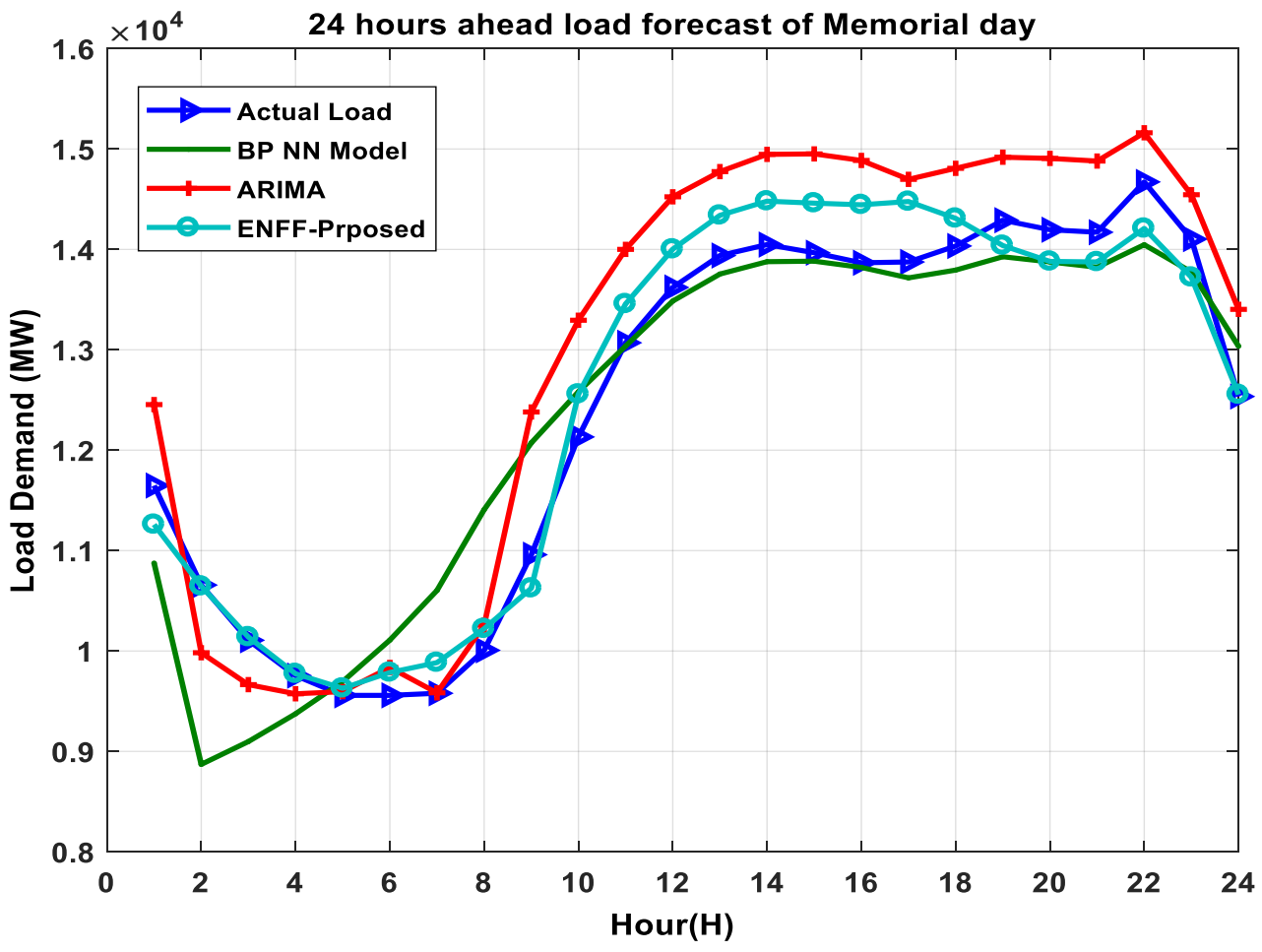

Fig. 7. The 24-hours ahead load forecast of Memorial Day.

## *B. Autoregressive Integrated Moving Average (ARIMA)*

ARIMA based forecast model, differenced series appearing in the forecasting equation and lags of the forecast errors are named as autoregressive and moving average respectively. The purpose of including a time series predictors in ensemble network is to achieve the diverse forecast output. In an ensemble network, each standalone predictor will explore the different forecast possibilities. As a result, the overall forecast accuracy of the proposed ensemble would be enhanced by integrating each predictor [25].

## *C. Forecast test case studies*

The historical highly correlated load data and the respective weather data are applied as inputs to the forecasting framework. Moreover, ARIMA and BPNN based forecast models are also applied to predict the load demand of anomalous days to compare the forecast results of the proposed ENFF forecast model with the identical simulation parameters.

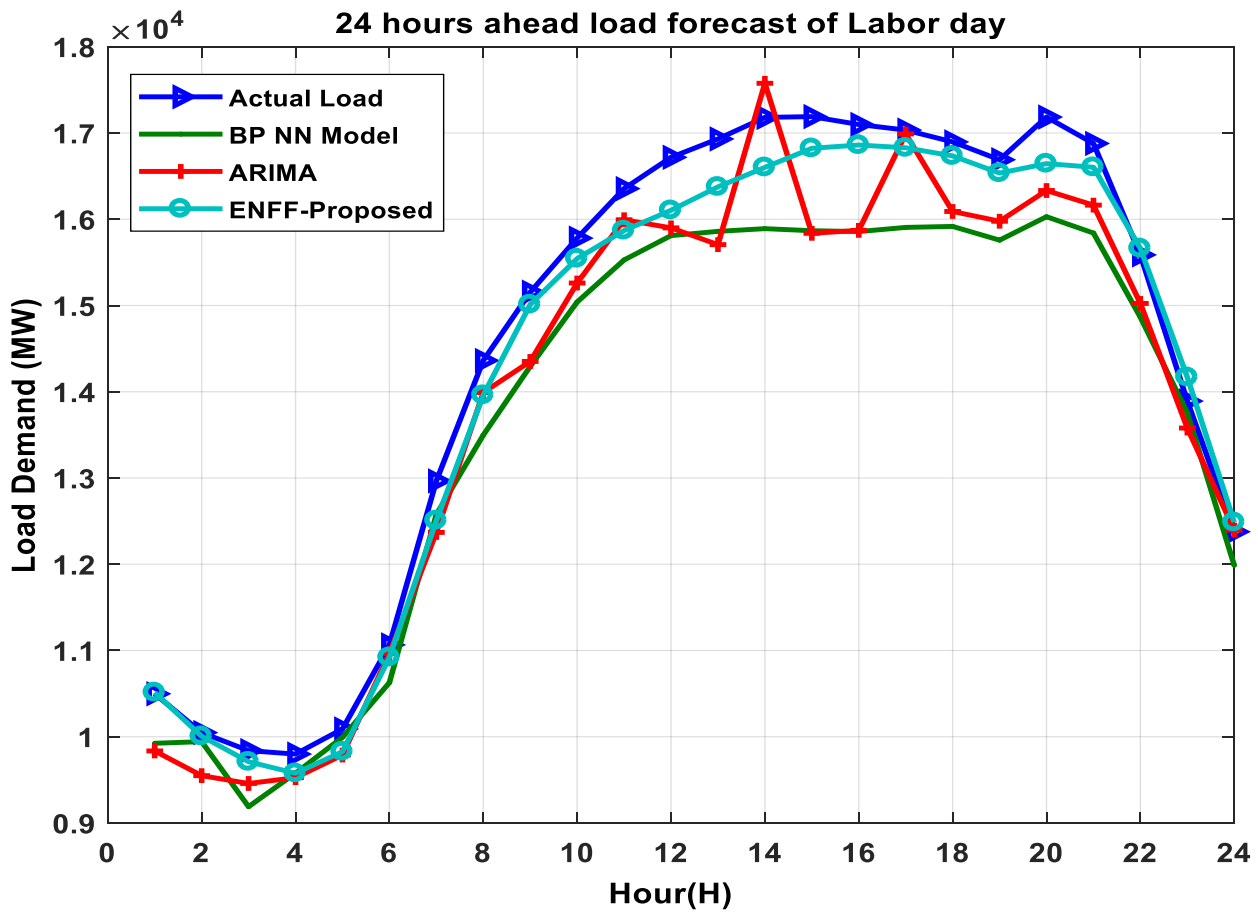

Fig. 8. The 24-hours ahead load forecast of Labor Day.

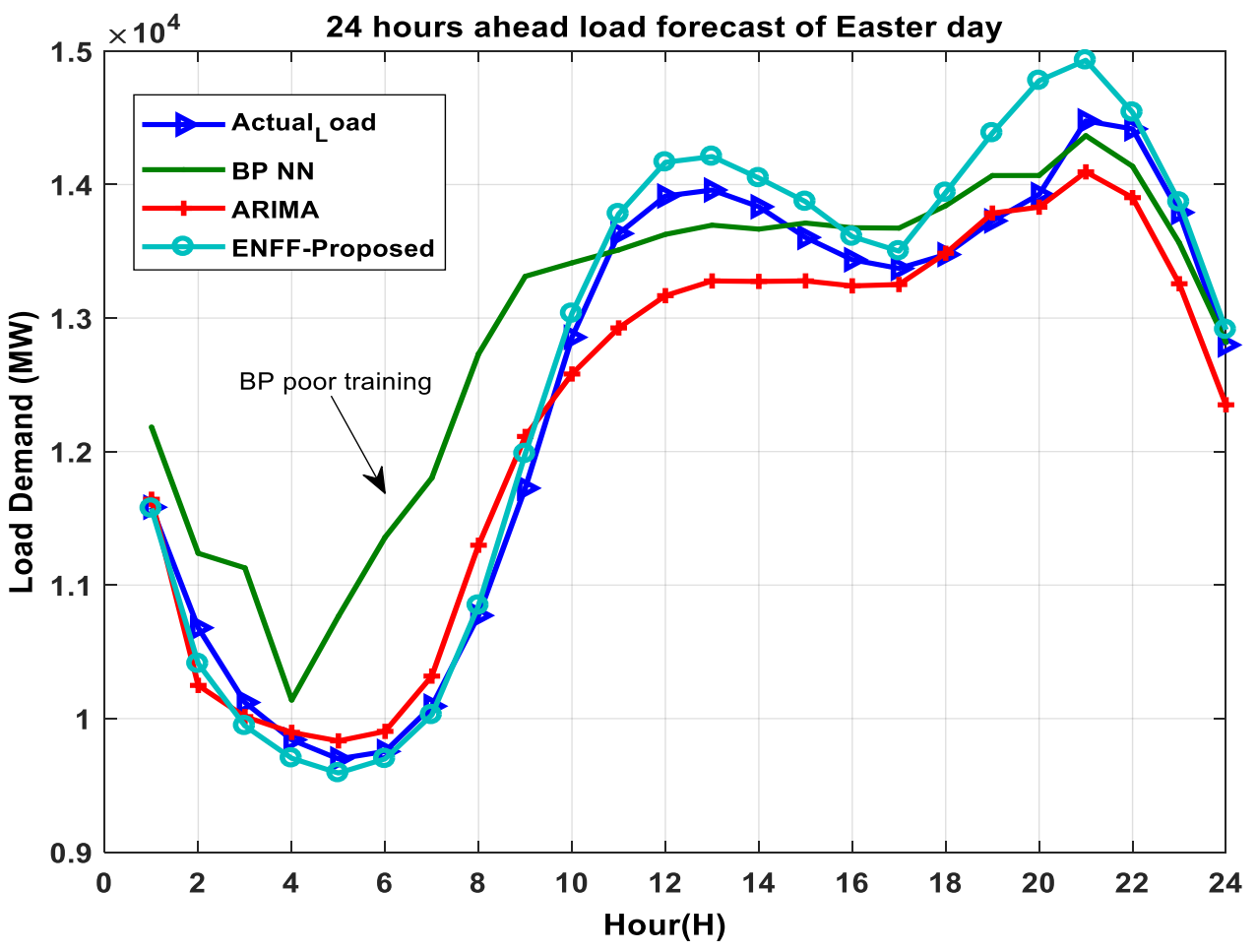

Fig. 9. The 24-hours ahead load forecast of Easter Day.

In order to predict the load demand on anomalous days, the network requires a longer data length with similar load patterns for proper training of the network. The reason being that few data patterns of previous anomalous days are available in the input database to properly train the network. Therefore, in this research, a five-year hourly load data and respective weather data from 2004 to 2008 are applied to the network for high forecasting accuracy.

The *x*-axis of the graph represents the hours of the anomalous day and the *y*-axis represents the load demand in megawatt. It can be observed from Fig. 6 that the proposed model produces less forecast error as compared to BPNN and ARIMA techniques. The forecast MAPE of ENFF, BPNN, and ARIMA are 1.85, 4.57, and 5.61%, respectively. Fig. 6 also depicts that the BPNN forecast model gives large forecast error from the 5-th to the 12-th hour of the day due to poor network training and less generalization capability over a large input training data. Conversely, the proposed forecast framework produced higher forecast accuracy due to better network training and higher generalization over a large input data length. Fig. 7 illustrates the comparison of the predicted load demand of the proposed ENFF along with ARIMA and BPNN on Memorial Day.

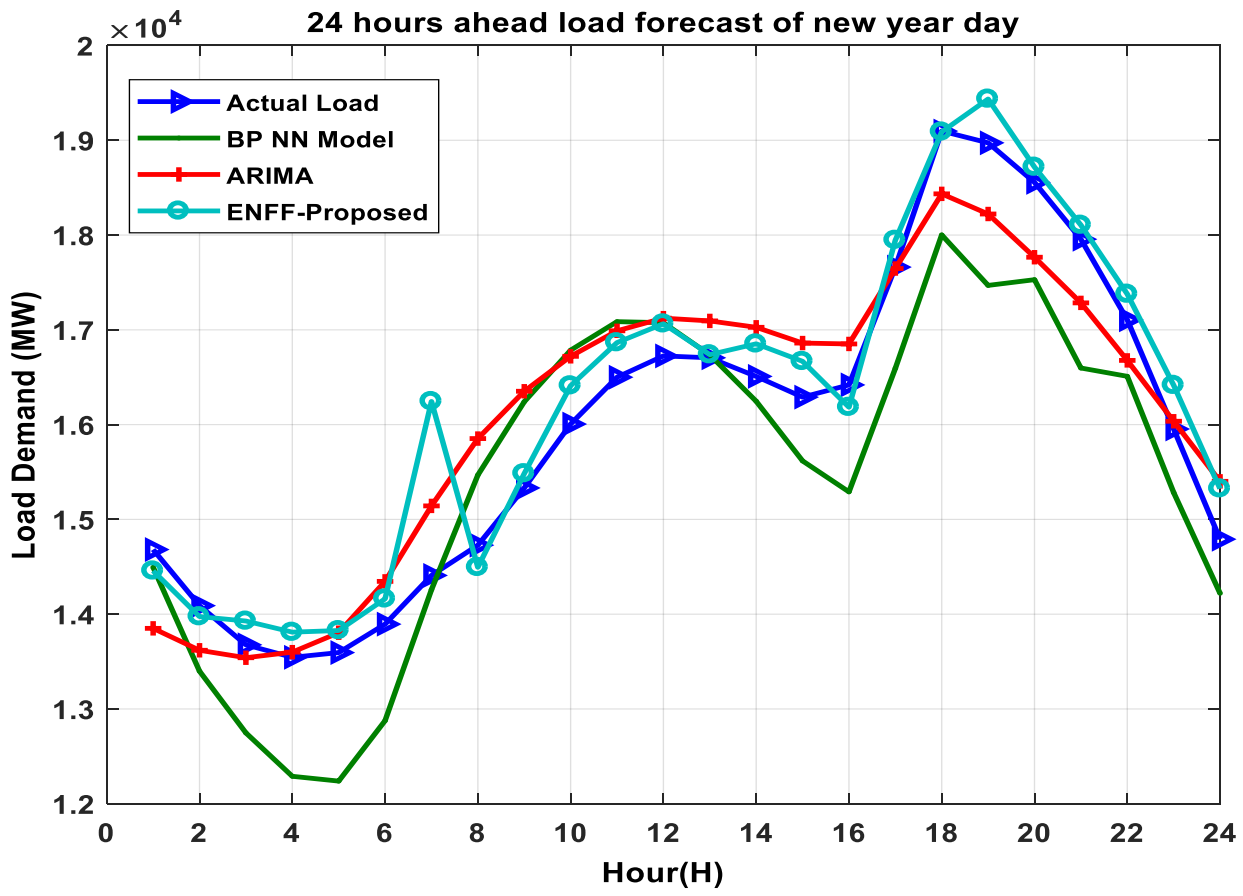

Fig. 10. The 24-hours ahead load forecast of New Year's Day.

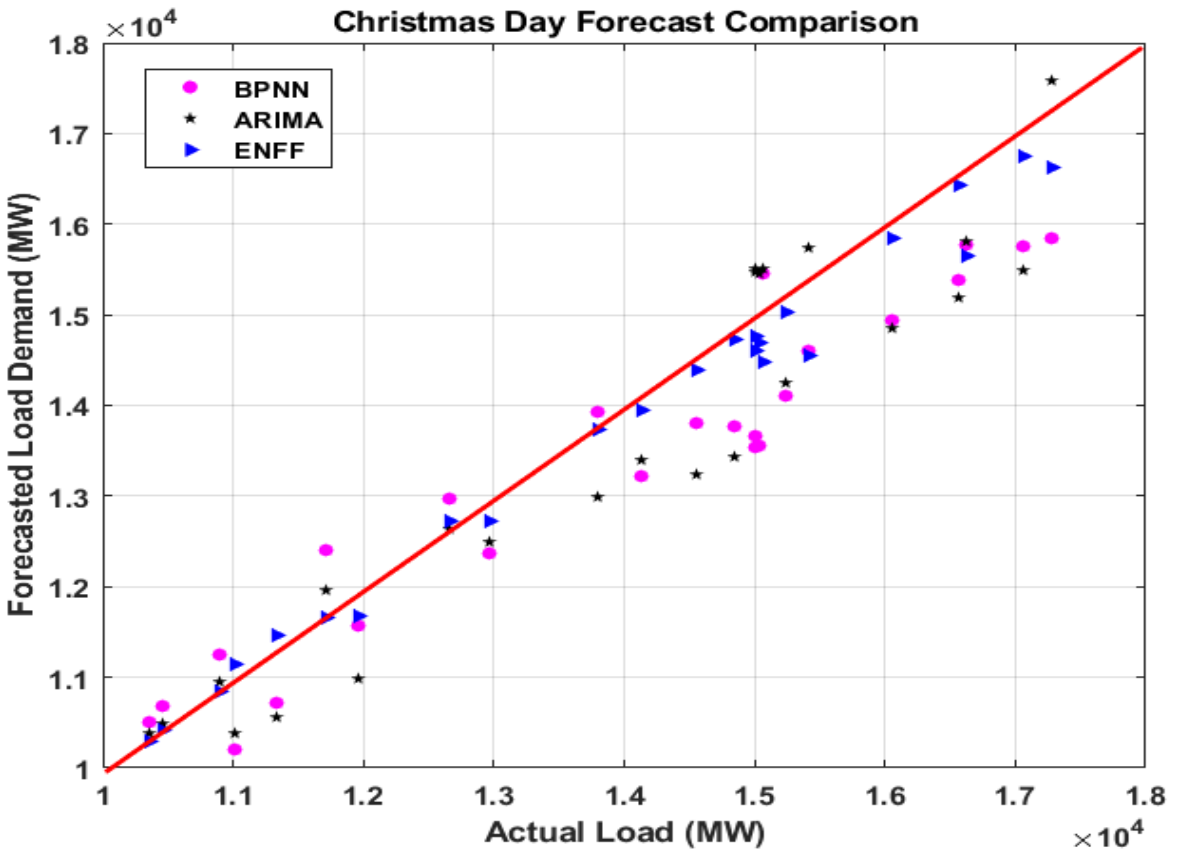

Fig. 11. Scatter plot comparison of predictors for Christmas Day.

The forecast MAPE of ENFF, BPNN, and ARIMA forecast models are 2.36%, 3.89%, and 4.51%,respectively. A similar forecast error trend is also observed in this case study, which shows that BPNN gives higher forecast error than ARIMA and ENFF. Fig. 8 also illustrates that the proposed framework predicted load demand closer to the actual load data of the Labor Day and also produced less MAPE; the forecast MAPE is 1.87, 3.90 and 5.22% for ENFF, BPNN and ARIMA forecast models, respectively. Figs. 9 and 10 represent the predicted load demand comparison of Easter Day and New Year's Day celebrated in the year 2009.

For Easter Day case study forecast results show that ENFF, ARIMA, and BPNN produced MAPE of 1.77%, 2.52% and 5.22%, respectively, and 2.15%, 3.27% and 4.96%, respectively, for the New Year's Day case study. Moreover, it also can be observed from the Easter Day and New Year's Day load forecast results that the BPNN model produces a higher error during the first ten hours of a day than the rest of the day load forecast due to insufficient training of the network and uncertainty. The ARIMA model gives higher forecast accuracy than the BPNN model and lower than the proposed model. The proposed ENFF forecast model produces higher forecast accuracy as it combines the output of multiple predictors in a framework.

Figs. 11 to 16 represent the scatter plot between predicted and actual load demand using the proposed ENFF, ARIMA and BPNN models for selected anomalous days.

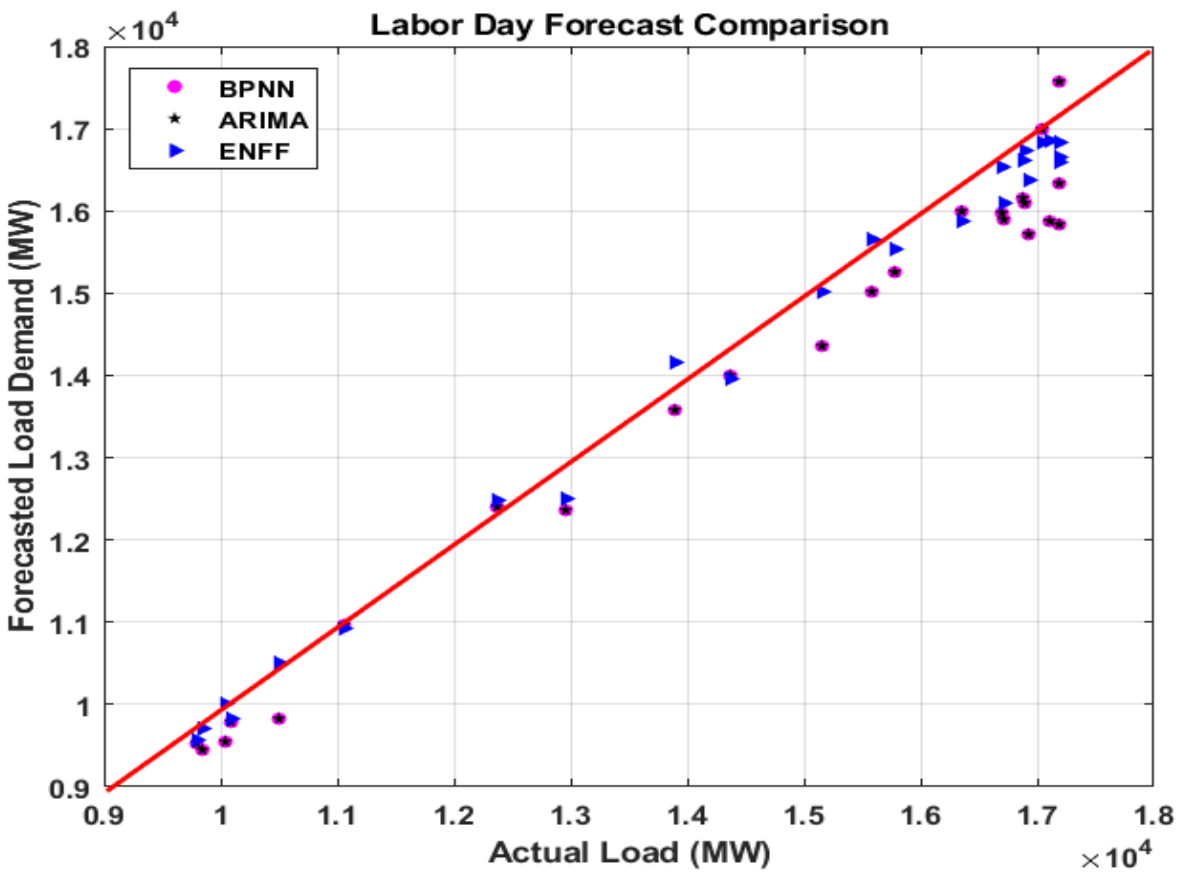

Fig. 12. Scatter plot comparison of predictors for Labor Day.

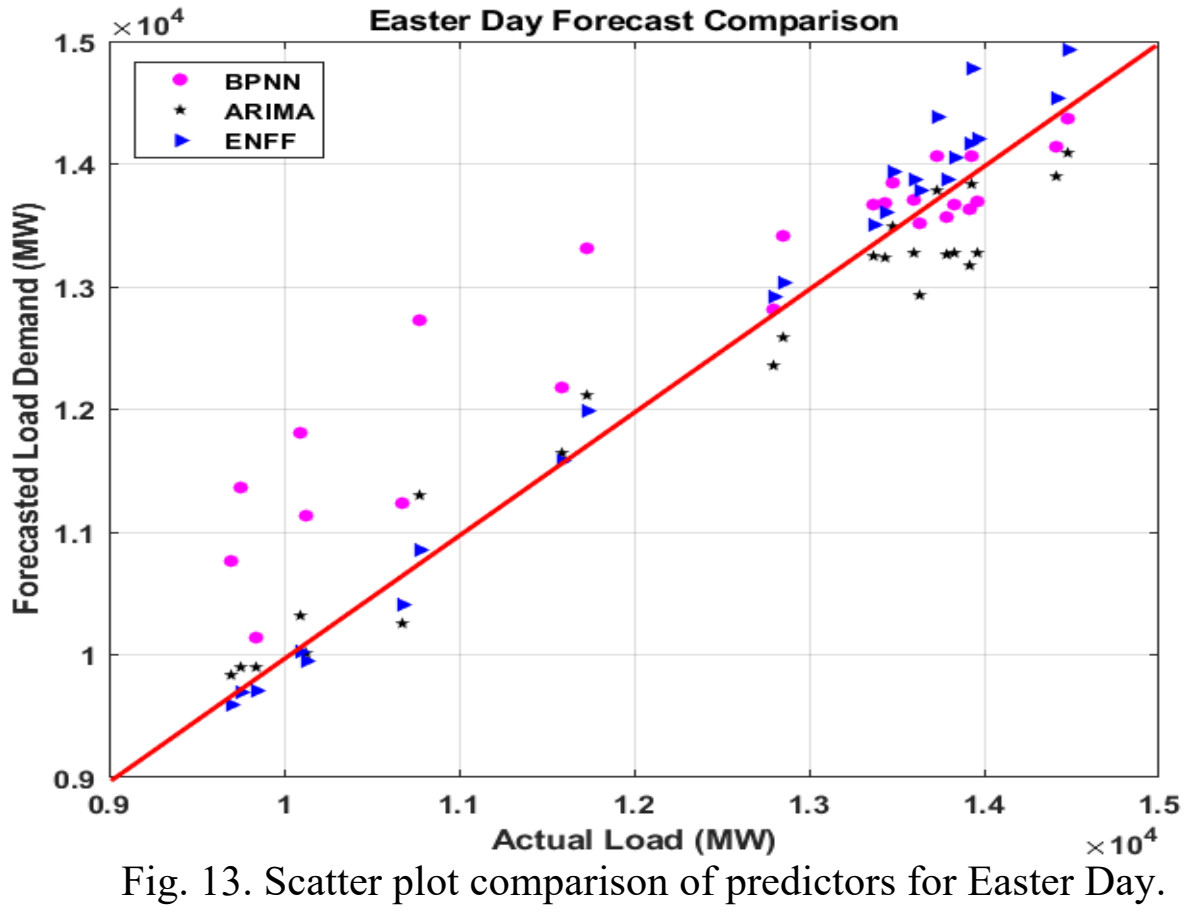

Fig. 13. Scatter plot comparison of predictors for Easter Day.

The *x*-axis depicts the actual load and the *y*-axis shows the predicted load demand. In Fig. 11, BPNN scatter points (predicted load) are apart from an ideal forecast line. It can be observed that BPNN either under or over forecasts the load demand for Christmas Day. Similarly, it can also be observed in Fig. 12 that the predicted load points by BPNN are dispersed and less close to the actual load in comparison to others. A similar forecast trend can be observed for the ARIMA based forecast model as shown in Figs. 11 and 12.

The results indicate that the BP algorithm is unable to train the network efficiently and gives poor performance uncertainty for the test day. Moreover, potentially it leads to poor training performance. As a result, it will increase the network forecast error. However, the proposed ENFF model scatter plots show that the predicted points are more concentrated to the ideal forecast line. This indicates that the predicted load demand is closer to actual load demand and produces higher forecast accuracy. A large dispersion of scatter points of BPNN and ARIMA models can be observed in Fig. 15. This large dispersion indicates that the BPNN and ARIMA models are not able to deal with uncertainty. Therefore, data uncertainty greatly affects the performance of NN based forecast and decreases the forecast accuracy. Moreover, ARIMA and BPNN based model produce large forecast errors during Easter Day test period as shown in Fig. 13. Overall, ENFF based forecast framework produces better forecast results than the BPNN and ARIMA. The predicted load demand using the ENFF model is closer to the actual demand. Especially, the ENFF model gives improved prediction accuracy and demonstrate the higher capability to deal with uncertainty in load that occurred during Easter Day.

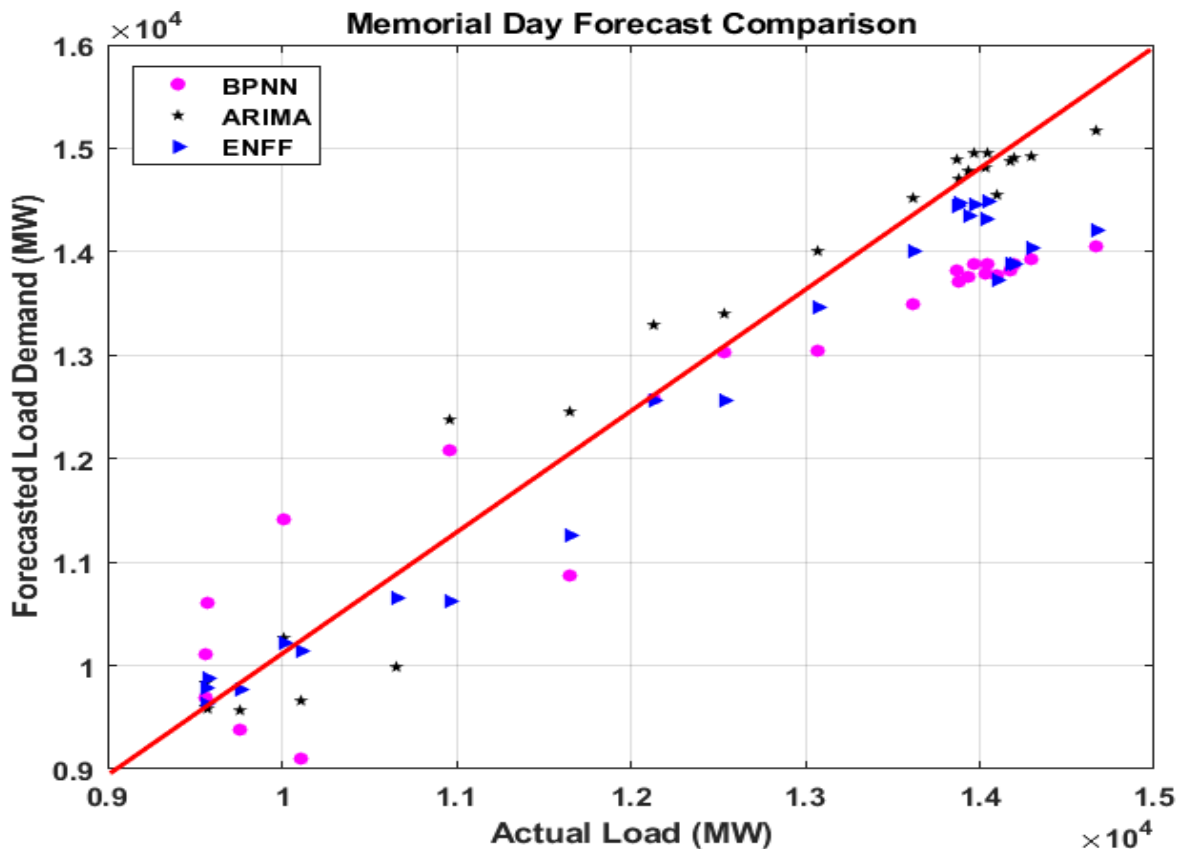

Fig. 14. Scatter plot comparison of predictors for Memorial Day.

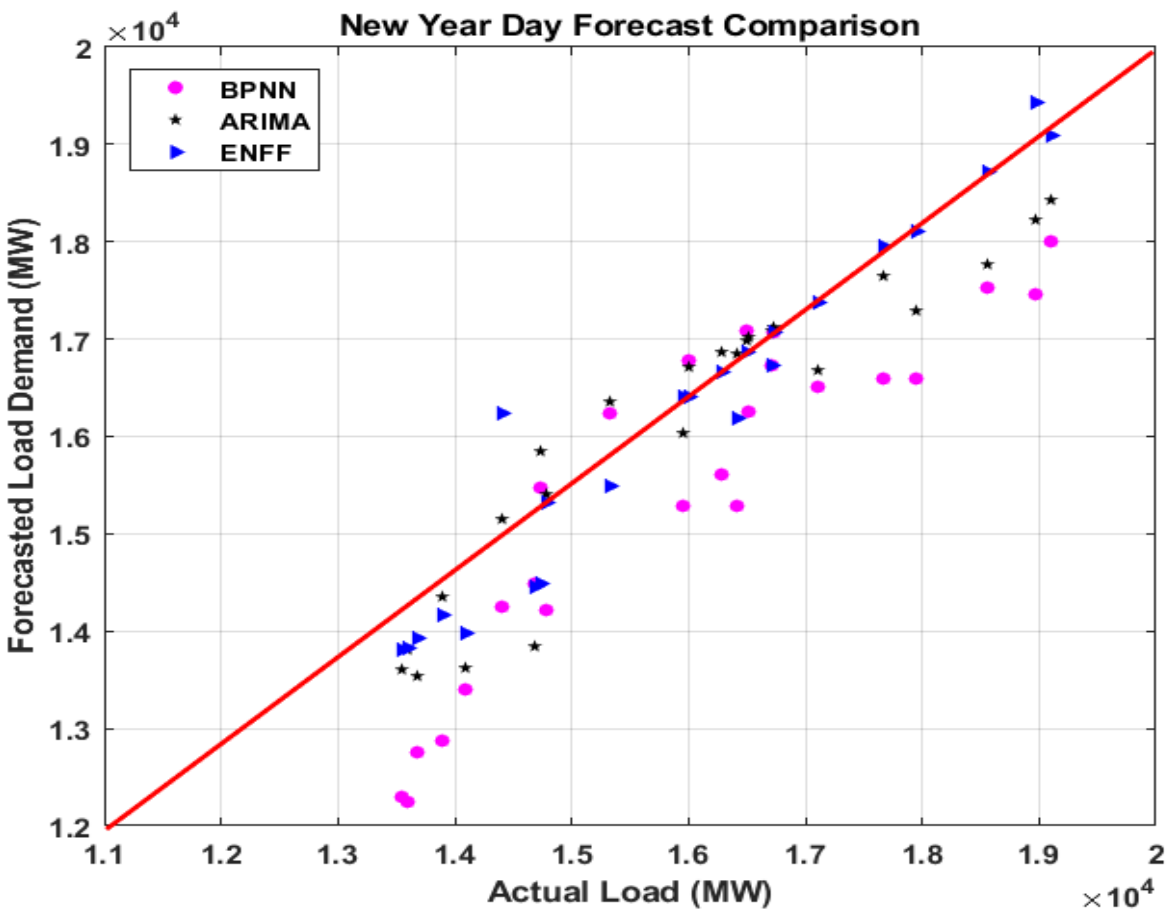

Fig. 15. Scatter plot comparison of predictors for New Year's Day.

### *D. Anomalous Day's Forecast MAPE Comparison of ENFF with BPNN and ARIMA*

The anomalous day's load forecast comparison of the proposed model and benchmark techniques is summarized in Fig. 16. It can beobserved that the proposed ENFF outperforms all anomalous day load forecasts by the benchmarks techniques. The comparative analysis of anomalous day's forecasts studies illustrates that the BPNN forecast model produces higher forecast MAPE than the ENFF and ARIMA. It can also be analyzed that BP technique gives the highest MAPE due to poor network training and lower generalization capability. Local minima problem arises in the BP training technique. As a result of the local minima problem, the network cannot be trained properly. Conversely, GPSO trained network has the capability to avoid the local minima problem. Forecast results show that the proposed GPSO based model gives higher forecast accuracy in all anomalous day case studies. It means that GPSO based NN forecast model can achieve a higher accuracy due to an efficient way to train the NN. It also proves that the GPSO based forecast model shows better generalization capability over a large training data of the forecast model as found in [36, 37].

## V. Conclusions

In this paper, an ensemble forecast framework (ENFF) is proposed with a systematic combination of three different predictors named Elman neural network (ELM), feedforward neural network (FNN) and radial basis function (RBF) neural network. Predictors are trained with GPSO and inputs are preprocessed using wavelet transformation to improve the prediction performance of predictors in the ensemble framework. The output of predictors is combined after removing the outliers using a trim aggregation technique.

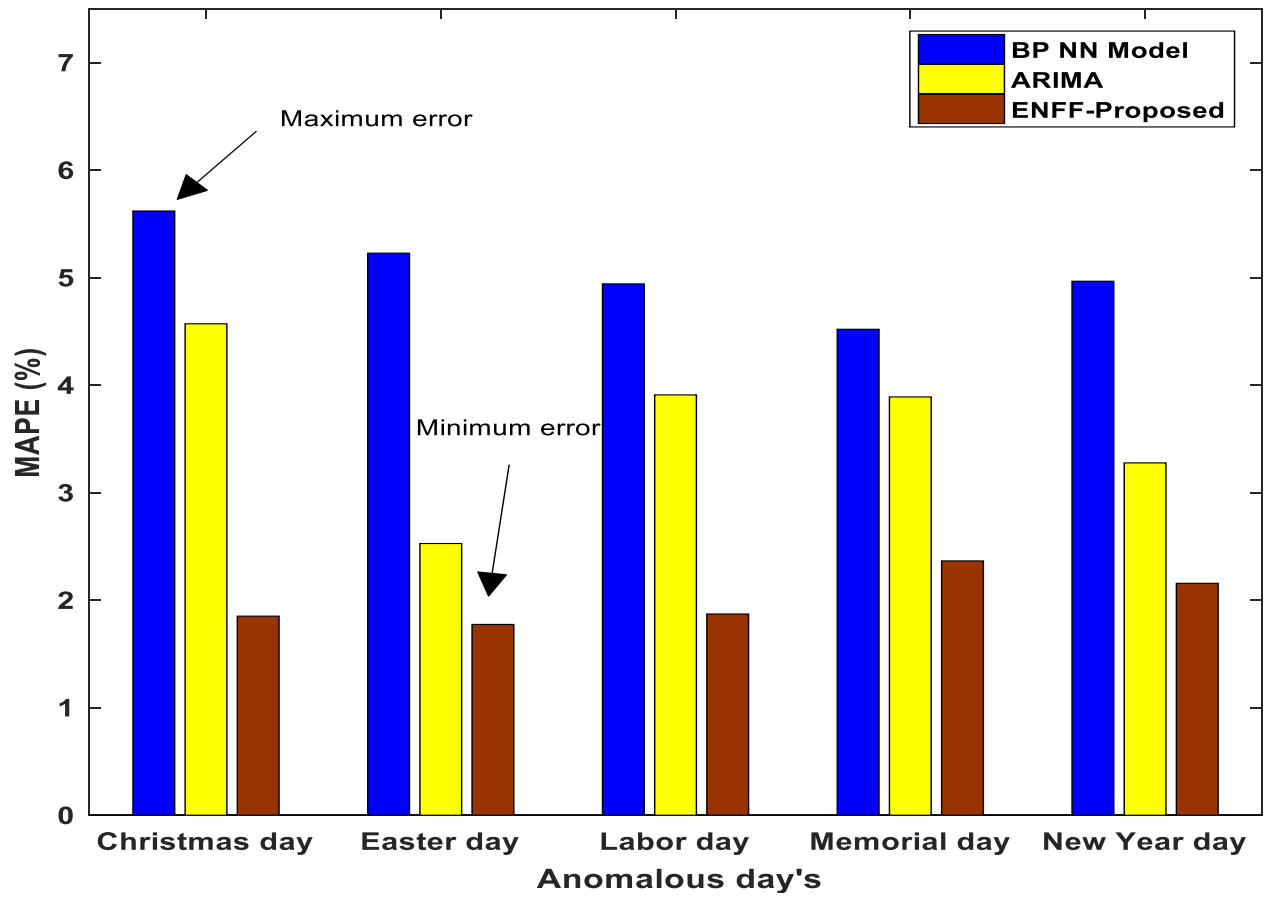

Fig. 16. Anomalous days load forecast error comparison.

Correlated historical load demand and meteorological variables are used as inputs to the multivariate ensemble forecast framework. The load forecast results for anomalous days celebrated in the year 2009 show that the ENFF model decreases the (percentage error) prediction error by a maximum of 52.09% in comparison with BPNN and ARIMA models. In addition, ENFF also produces less forecast MAPE up to 1.77% in comparison with BPNN and ARIMA based models. The ENFF gives minimum forecast error due to better training of the network and capability to avoid the local minima problem, unlike the BP algorithm. It can be concluded that prediction accuracy can be improved by combining multiple predictors in an ensemble framework. In the future, the proposed ensemble framework will be used for other forecasting applications such as PV output power, wind speed and natural gas demand forecasting.

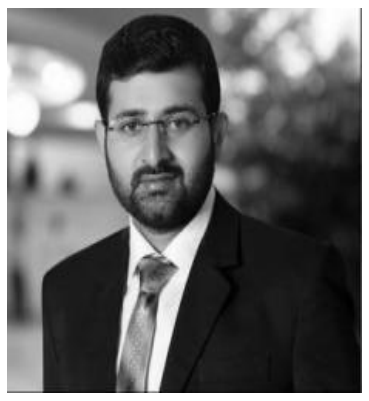

**Muhammad Qamr Raza (M'2012)** graduting with Ph.D. degree in electrical engineering from The University of Queensland, Brisbane, Australia in 2018. He received the B.Sc. degree in electrical and electronic engineering (EEE) from COMSATS IIT, Pakistan and M.Sc. by research degree in EEE from Universiti Teknologi PETRONAS (UTP), Malaysia, in in 2010 and 2014 respectivly.

He is serving as associate editor of International Journal of Electrical Engineering Education (Published from UK since 1963). He also served as reviewer for several reputed journals such as IEEE Transaction on Smart Grid, Power System and IEEE conferences. His research interests includes machine learning, smart buildings, electrical load, price and PV output power forecasting.

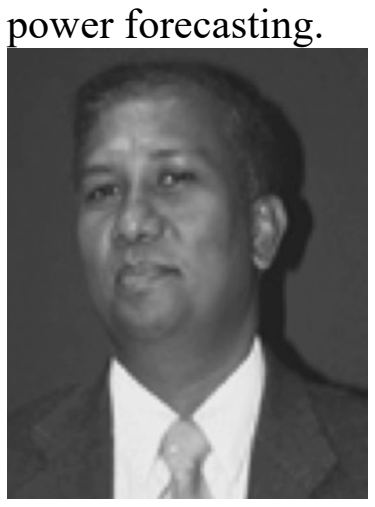

**Nadarajah Mithulananthan (SM'10)** received the Ph.D. degree in electrical and computer engineering from the University of Waterloo, Waterloo, ON, Canada, in 2002. He was an Electrical Engineer at Generation Planning Branch of Ceylon Electricity Board, Sri Lanka, and a Researcher at Chulalongkorn University, Bangkok, Thailand. He also served as an Associate Professor at the Asian Institute of Technology, Bangkok. Currently, he is with the School of Information Technology and Electrical Engineering at the University of Queensland, Brisbane, Australia. His main research interests are renewable energy integration and grid impact of distributed generation, electric-vehicle charging, and energy-storage systems.

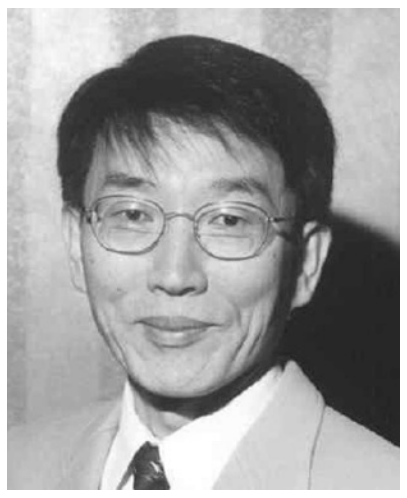

**Kwang Y. Lee (F'01)** received the B.S. degree from Seoul National University, Seoul, Korea, in 1964, and the M.S. degree from North Dakota State University, Fargo, in 1968, both in electrical engineering, and the Ph.D. degree in system science from Michigan State University, East Lansing, in 1971.

He is currently a Professor and Chair of electrical and computer engineering with Michigan State University, Oregon State University, Newport, University of Houston, Houston, TX, the Pennsylvania State University, University Park, and Baylor University, Waco, TX. His current research interests include power system control, operation, planning, and intelligent system applications to power systems. Dr. Lee is an Editor of the IEEE Transactions on Energy Conversion. He was an Associate Editor of the IEEE Transactions on Neural Networks. He is also a Registered Professional Engineer.

**Jiaming Li** is associated with Data61 energy grooup, Commonwealth Scientific and Industrial Research Organization (CSIRO), Sydney, Australia. Author further details are not available at time of publicaiton.